\documentclass[runningheads]{llncs}
\usepackage{graphicx}
\usepackage{times}
\usepackage{epsfig}
\usepackage{graphicx}
\usepackage{amsmath}
\usepackage{amssymb}
\usepackage{colortbl}
\usepackage[ruled,vlined]{algorithm2e}
\begin{document}
\title{Vascular surface segmentation for intracranial aneurysm isolation and quantification}
\titlerunning{Vascular surface segmentation}
%

\author{\v{Z}iga Bizjak\inst{1} \and
Bo\v{s}tjan Likar\inst{1} \and
Franjo Pernu\v{s}\inst{1} \and \v{Z}iga \v{S}piclin\inst{1}}

\authorrunning{Bizjak et al.}
\institute{${}^1$University of Ljubljana, Faculty of Electrical Engineering\\Tr\v{z}a\v{s}ka cesta 25, 1000 Ljubljana, Slovenia\\ \email{ziga.bizjak@fe.uni-lj.si}\\
}

\maketitle
\begin{abstract}Predicting rupture risk and deciding on optimal treatment plan for intracranial aneurysms (IAs) is possible by quantification of their size and shape. For this purpose the IA has to be isolated from 3D angiogram. State-of-the-art methods perform IA isolation by encoding neurosurgeon’s intuition about former non-dilated vessel anatomy through principled approaches like fitting a cutting plane to vasculature surface, using Gaussian curvature and vessel centerline distance constraints, by deformable contours or graph cuts guided by the curvature or restricted by Voronoi surface decomposition and similar. However, the large variability of IAs and their parent vasculature configurations often leads to failure or non-intuitive isolation. Manual corrections are thus required, but suffer from poor reproducibility. In this paper, we aim to increase the accuracy, robustness and reproducibility of IA isolation through two stage deep learning based segmentation of vascular surface. The surface was represented by local patches in form of point clouds, which were fed into first stage multilayer neural network (MNN) to obtain descriptors invariant to point ordering, rotation and scale. Binary classifier as second stage MNN was used to isolate surface belonging to the IA. Method validation was based on 57 DSA, 28 CTA and 5 MRA images, where cross-validation showed high segmentation sensitivity of 0.985, a substantial improvement over 0.830 obtained for the state-of-the-art method on the same datasets. Visual analysis of IA isolation and its high accuracy and reliability consistent across CTA and DSA scans confirmed the clinical applicability of proposed method.
\end{abstract}

\section{Introduction}
Intracranial aneurysms (IAs) are abnormal bulges mainly located at vessel bifurcations and arising from weakened vessel wall. 
Despite being highly prevalent ($\sim$3.2\% of population), most IAs do not rupture throughout patient's lifetime and are not associated to any symptoms. In case of rupture, however, the subsequent subarachnoid hemorrhage (SAH) has 50\% fatality rate. Imaging studies have shown that aneurysm size and shape are one of the key factors for prediction rupture risk and deciding on optimal treatment plan for unruptured IAs \cite{ishibashi_toshihiro_unruptured_2009}. For instance, in ELAPSS score~\cite{backes_elapss_2017} the IA size can amount to 55\% of the final score, whereas ELAPSS also takes into consideration earlier SAH, location of aneurysm, age, population and shape of aneurysm. 

Clinical treatment of small and medium IAs is yet not well defined. Note that size groups of IAs are small (0-4.9 mm in diameter), medium (5 to 9.9 mm), large (10 to 25 mm) and giant ($>$25 mm)~\cite{ishibashi_toshihiro_unruptured_2009}. 
Studies have shown that risk of complications during procedure is larger than risk of spontaneous rupture \cite{brinjikji_risk_2016,ishibashi_toshihiro_unruptured_2009,brown_unruptured_2014,wiebers_unruptured_2003,chien_unruptured_2019}. There are many cases, where risk of spontaneous rupture is less likely than the risk due to treatment and where follow-up imaging is considered to be the best course of treatment. Follow-up imaging may involve several modalities like DSA, CTA and MRA. Therefore, an automatic and reproducible modality-independent process of aneurysm size and shape measurement is essential to provide best achievable treatment for patient and to correctly measure change of aneurysm morphology between consecutive imaging sessions. 

Aneurysm size and shape quantification is possible through its segmentation and isolation from parent vessels. Most neurosurgeons still use manual methods to isolate IAs \cite{cardenes2013performance}. 
Due to high intra- and inter-rater variability of manual aneurysm segmentation, and the time-consuming and cumbersome 3D image manipulation, there is a need for fast, accurate and consistent computer-assisted segmentation methods. 

In this paper we propose surface point classification (SPC) method to segment and isolate the IA. Once the point cloud is extracted from 3D image, for instance by simple interactive thresholding, the procedure to isolate IAs remains the same for all modalities. Our segmentation is based on the point cloud data and is thus modality-independent. This was successfully verified on a total of 100 IA cases, with 57 DSA, 28 CTAs and 5 MRA scans. The proposed method achieved consistent performance across different image modalities and also improved segmentation over the state-of-the-art.







\section{Related work}

\begin{figure}[!t]
	\begin{center}
		\includegraphics[width=4.5in]{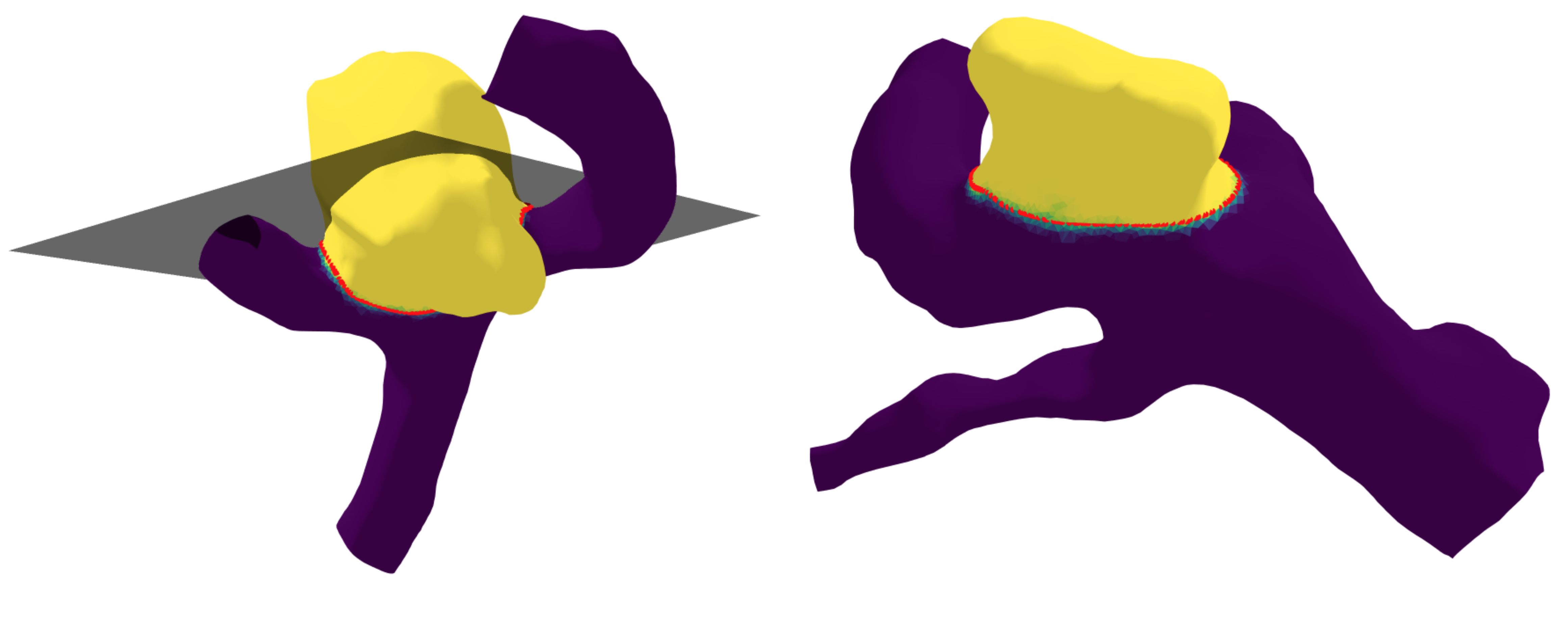}
	\end{center}
	
	\caption{\small Two examples of aneurysm segmentation. Current state-of-the-art-method (ACP) fails at isolate aneurysm with plane (\textit{grey surface on the left}), while our SPC method (\textit{yellow-colored surface}) is very close to reference manual segmentation (\textit{red curve}). }
	\label{model_output}
\end{figure}

An approach mimicking manual IA isolation is by positioning a cutting plane. In past clinical practice the cutting plane was chosen manually with the help of the 3D visualization software~\cite{ma2004three}. Jerman et al.~\cite{jerman2019automated} developed method that automatically positions the cutting plane (ACP). This method works well on most of the aneurysms and is considered as current state-of-the-art. However, if the aneurysm is small or if the aneurysm is blended with surrounding vessels the ACP may fail. Furthermore, if the shape of IA's neck does not lie in a plane, the method may again fail to isolate aneurysm correctly as shown in Figure~\ref{model_output}. 

Non-planar curve for separation of aneurysm from the surrounding vessels seem a better option. Ruben et al.~\cite{cardenes2011automatic} used automatic approach based on the computation of a minimal path along a scalar field obtained on the vessel surface. In order to assure correct topology of the aneurysms bulb, the neck computation was constrained with a region defined by surface Voronoi diagram. The method was evaluated on 26 real cases against manual aneurysm isolation using cutting plane. This method works well for standard cases, but can fail if the configuration of the arteries is complex. Marina et al.~\cite{piccinelli2012automatic} proposed a method also based on Voronoi diagram with the combination of so-called tube function. The method was evaluated on 30 cases. In 20 cases, the manual and computed separation curves were considered as being very similar and both acceptable. In five cases manual curve was better then the computed, but still acceptable. In five cases the method did not produce similar curves compared to manual ones. Hence, robustness remains an issue with non-planar separation curves.

Sylivia et al.~\cite{saalfeld2018semiautomatic} proposed a semi-automatic separation curve reconstruction algorithm for automatic extraction of morphological parameters. The first drawback of this approach is that it requires a pronounced aneurysm neck to work properly. The second drawback is that the user has to click on the aneurysm to initialize the method's parameters. Lawonn et al.~\cite{lawonn2019geometric} proposed a geometric optimization approach for the detection and segmentation of multiple aneurysms in two stages. First, a set of aneurysm candidate regions was identified by segmenting regions of the vessels. Second stage was a classifier, that identified the actual aneurysms among the candidates. While the method was capable of detecting aneurysm location automatically, the user still had to use the brush tool to get correct aneurysm neck curve. To address this problem they proposed a smoothing algorithm based on ostium curve extraction to improve upon previous results. While smoothing algorithm worked fine, additional manual user input was still necessary for some aneurysm cases.

All of the aforementioned methods mostly provide an acceptable result for large saccular IAs that have a well defined aneurysm neck (the cross section of the inlet is significantly smaller than cross-section of the aneurysm). For other aneurysm size and shapes most of these methods fails to deliver correct result. To our knowledge, there is still no fully automatic method that would work well for IAs of various sizes and shapes, regardless of the configuration of surrounding vessels. 

\section{Materials and methods}

Current clinical detection and measurements of IAs is performed on angiographic images like DSA, CTA and MRA (section~\ref{cases}). Procedures used to extract 3D surface mesh from angiographic images are explained in section~\ref{preprocesing}, data augmentation and training process  in~\ref{training} and the inference on clinical images in~\ref{inference}.

\begin{figure}[]
	\begin{center}
		\includegraphics[width=4.7in]{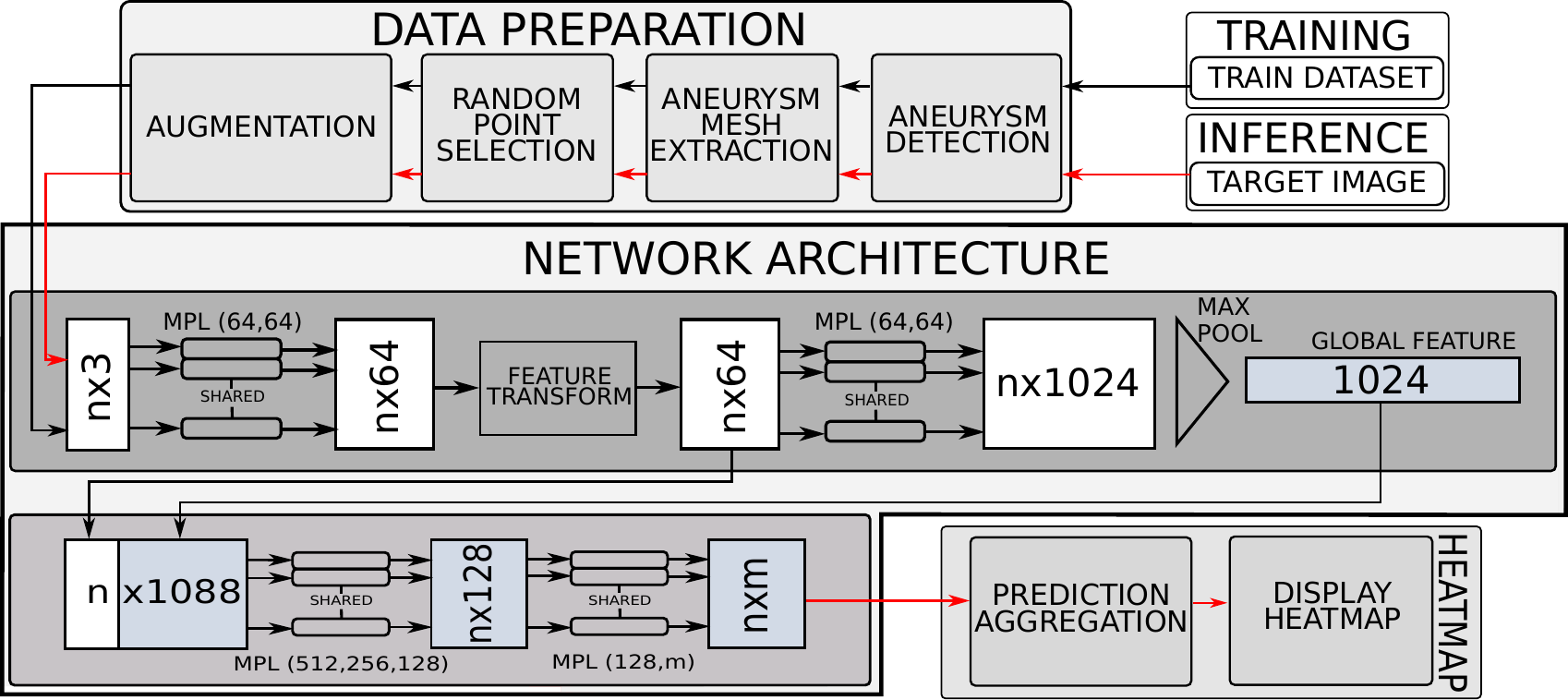}
	\end{center}
	
	\caption{\small Flowchart of the proposed aneurysm segmentation method.}
	\label{PointNet}
\end{figure}

\subsection{Case Information}\label{cases}
This study was performed with the approval of the institutional review board. A total of 100 patient angiographic 3D scans, including 57 DSA images, 28 CTA images and 5 MRA images, of the cerebrovascular region were obtained for the purpose of this study. 
All images were acquired at University Medical Centre Ljubljana
using standard imaging protocols used in clinical routine. 
Per patient incidence of IAs in our dataset was from zero to three. We found a total of 100 IAs, from which 27 are considered small ($<5$ mm), 49 medium sized ($5<$ size $<10$ mm) and 24 are large ($>25$ mm). The median diameter of the aneurysms was 6.99 mm. 

\subsection{Mesh extraction from 3D angiographic image}\label{preprocesing}

First step in extracting the IA mesh from 3D angiographic image is IA detection. The search for IA is mainly still performed by skilled surgeon without any special computer-assisted tools. Many algorithms for automatic aneurysm detection were previously proposed \cite{jerman2017aneurysm,zhou2019intracranial,jin2019fully,duan2019automatic}, but all have the difficulty of presenting with to many false positives or are not able to detect small aneurysms. For our purposes detection was carried out manually using visual image inspection.

Following detection, a region of interest (ROI) containing the aneurysm and surrounding vessels was determined. Segmentation of vascular structures was performed by using interactive thresholding, followed by the application of marching cubes and smooth non-shrinking algorithms \cite{larrabide2011three,cebral2001medical}. In this way, the corresponding 3D surface mesh of aneurysm and surrounding vessels was reconstructed. The following procedures ware performed using image analysis software (RealGUIDE Software version 5.0; 3DIEMME Srl). Example mesh containing an aneurysm and surrounding vessels is shown in Figure~\ref{model_output}.

\subsection{Training}\label{training}

On every input mesh the aneurysm was manually isolated from surrounding vessels by skilled neurosurgeon by drawing a closed surface curve (see Figure \ref{model_output} for examples of manual reference curve for IA isolation). Aneurysm points were labeled according to the neurosurgeon's segmentation. Surgeon was not restricted to the specific shape of aneurysm neck. The obtained manual IA segmentation isolation were used for training and validation of multi-layer neural network (MNN) model~\cite{qi2017pointnet}.

To train MNN model, with network architecture as presented on Figure~\ref{PointNet}, that is robust to point ordering, rotations and scaling of input data we replicated each mesh \textit{n} times (\textit{n}=number of replications). Rotations were randomly applied with angles from 0 to 360 degrees in all three axes. Scaling factor was randomly set between values 0.5 and 1. From each augmented mesh we randomly picked \textit{m} points (\textit{m}=number of points) and saved them as point cloud. Augmented point clouds of one aneurysm appeared only in train, test or validation set.



\subsection{Inference - using trained model for isolation}\label{inference}


To further suppress MNN model's sensitivity to rotation, we applied a so-called multiple partial voting algorithm. Similar to the augmentation process before training, we used the same augmentation procedures for the target mesh until each point of the target mesh was used at least 20 times, each time with different rotation. For every augmentated instance we used \textit{m} random points from target mesh. Each augmented point cloud contributed one vote to segmentation for each point. The sum of all votes was divided by number of votes for each point. The resulting heatmaps represented the output segmentation, which was compared to the other methods as shown on Figure~\ref{results}. 

To determine which points on heatmap form the aneurysm and which the surrounding vessels we used simple thresholding at 0.5. We also verified our decision on such threshold by plotting the ROC (receiver operating characteristic) curve.  

\section{Experiment}

The training of MNN segmentation model was executed only once. As train dataset we used 70\% of all available data, i.e. 70 aneurysm meshes. Each mesh was reproduced with different rotation about 100 times (\textit{n}=100) and 3000 points (\textit{m}=3000) were randomly chosen after each rotation. For test dataset we used 25\% of available data, i.e. 25 aneurysm meshes, which were also rotated \textit{n} times and \textit{m} points were chosen. We used 5\% of data or 5 aneurysm meshes for validation purposes. Those aneurysm meshes were used to create a valid heatmap. The number of epochs was 100 and the learning rate was 0.001 for fist 20 epochs, 0.0005 for epochs between 21-40, 0.00025 for epochs between 40-60, 0.000125 for epochs between 61-80 and 0.000063 for the rest of the epochs. Batch size was set to 16.

To fairly compare our results with the state-of-the-art method proposed by Jerman et al.~\cite{jerman2019automated} we also calculated SPC with plane (SPCP) as follows: using least-square fit a cutting plane was positioned onto border points between aneurysm and vessel.

The model was trained on the Linux based computer with 32 GB RAM, Intel Core 8700K 6 Core 4,7 GHz processor and 11GB NVIDIA Graphic Card.

\section{Results}

\begin{figure}[!t]
	\begin{center}
		\includegraphics[width=4.7in]{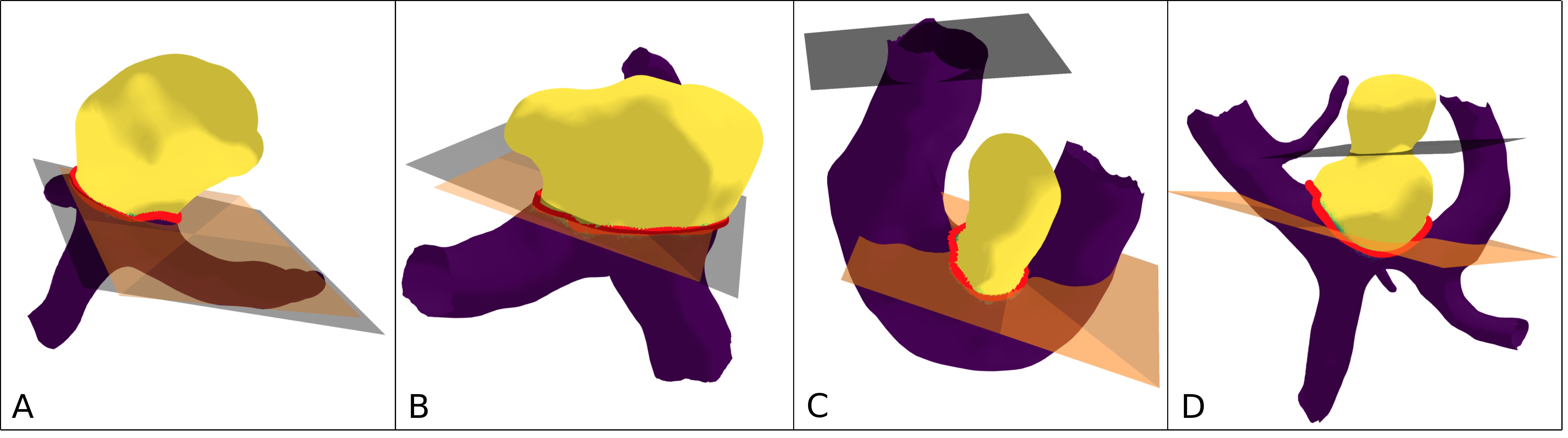}
	\end{center}
	
	\caption{\small Images \textbf{A-D} depict four examples of our isolation compared to gold standard and previous work. Red line represents gold standard aneurysm neck, grey plane represents the ouput of ACP algorithm, orange plane was fitted from boarder points of SPC. Yellow surface is our result of aneurysm isolation from vessels. Dark purple surface is predicted vessel surface.}
	\label{results}
\end{figure}

After training our model on 70 meshes, we tested the performance on all datasets. Point-wise predictions were rounded to the nearest integer (0 or 1) for all \textit{n} rotation of each mesh, the aggregated heatmap values of each target mesh were normalized to interval [0, 1]. For each mesh we computed true positive rate (TPR; sensitivity). Table~\ref{table-res} shows sensitivity evaluation of all data subsets. Median sensitivity on learning dataset was 0.985, 0.983 on test dataset and 0.994 on validation dataset. Overall median sensitivity was 0.985. If we fit plane to the SPC model result we achieve median sensitivity 0.938 on training dataset, 0.924 on test dataset and 0.947 on validation dataset. Overall median sensitivity was 0.938. Current state-of-the-art method achieved accuracy of 0.830 on training, 0.775 on testing and 0.846 on validation dataset, with median sensitivity of 0.830. State-of-the-art method failed to isolate aneurysm from the vessels in 8\% of all cases, while our proposed method succeeded in accurately isolating aneurysm from the vessel across all cases. The distribution of sensitivity for all three tested methods is presented in box plots on Figure~\ref{results}.

\begin{figure}[!t]
	\begin{center}
		\includegraphics[width=4.5in]{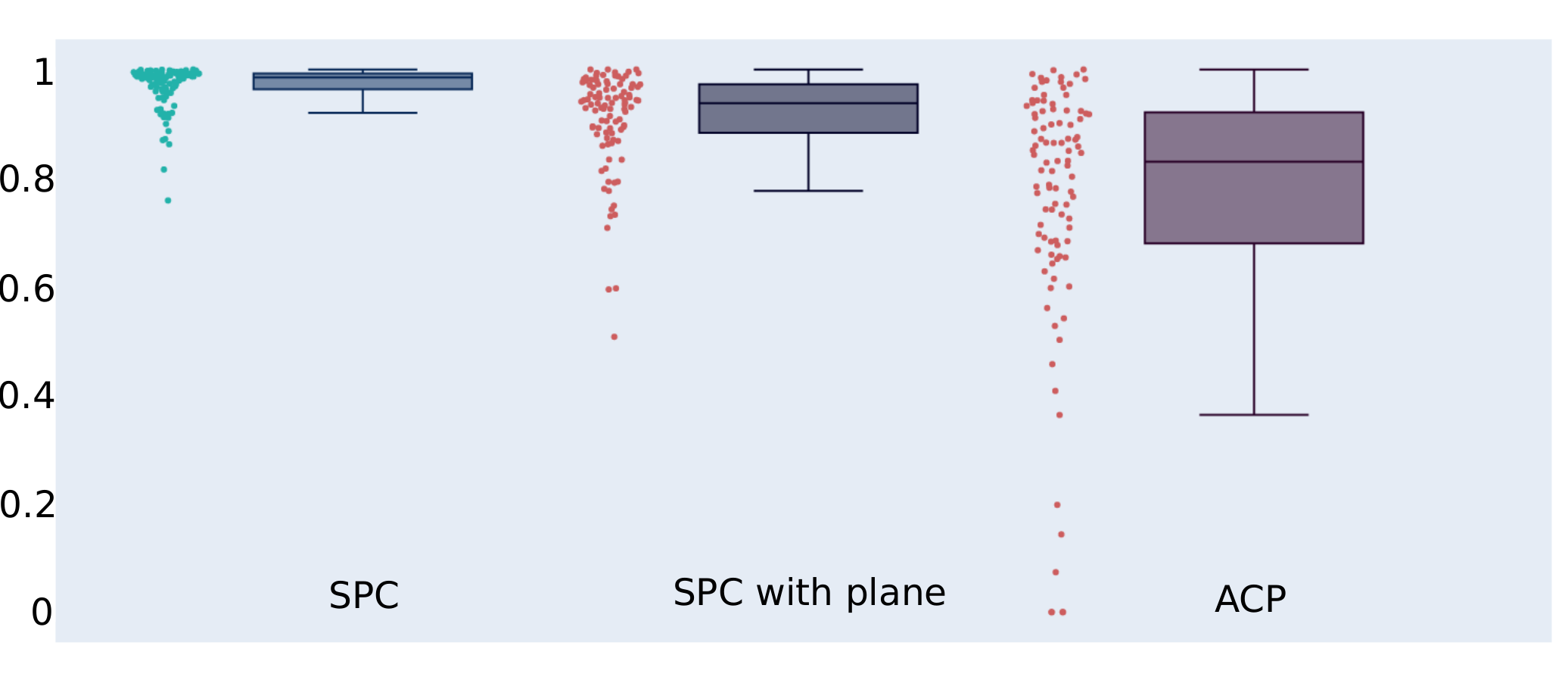}
	\end{center}
	
	\caption{\small Point-jitter and box plots of sensitivity values for all meshes for three tested methods.}
	\label{box_plots}
\end{figure}

Average time to segment and isolate one aneurysm from the vessels with the current state-of-the-art method was approximately 11 minutes. On the other hand, the isolation of using novel method executed in under 23 seconds per input image.

\begin{table}[]
\caption{Aneurysm segmentation evaluation across all 100 datasets and across the training, testing and validation cases.\label{table-res}}
\begin{tabular}{c|c|c|c|c|c|c|c|c|c|c|c|c}
\cline{2-13}
                                 & \multicolumn{3}{c|}{\textbf{ALL}} & \multicolumn{3}{c|}{\textbf{TRAIN}} & \multicolumn{3}{c|}{\textbf{TEST}} & \multicolumn{3}{c|}{\textbf{VAL}}                   \\ \hline 
\multicolumn{1}{|l|}{\textbf{\textbf{Method}}}     & \textbf{SPC}    & \textbf{SPCP}   & \textbf{ACP}    & \textbf{SPC}     & \textbf{SPCP}    & \textbf{ACP}    & \textbf{SCP}     & \textbf{SPCP}   & \textbf{ACP}    & \textbf{SCP}   & \textbf{SPCP}  & \multicolumn{1}{l|}{\textbf{ACP}}   \\ \hline \hline 
\multicolumn{1}{|l|}{min TPR}    & \textbf{0.759}  & 0.507  & 0      & \textbf{0.759}   & 0.507   & 0      & \textbf{0.918}   & 0.730  & 0.074  & \textbf{0.982 }& 0.892 & \multicolumn{1}{l|}{0.709} \\ \hline
\multicolumn{1}{|l|}{max TPR}    & 1      & 1      & 1      & 1       & 1       & 1      & 1       & 1      & 0.985  & \textbf{0.999} & 0.972 & \multicolumn{1}{l|}{0.972} \\ \hline
\multicolumn{1}{|l|}{median TPR} & \textbf{0.985 } & 0.938  & 0.830  & \textbf{0.985}   & 0.938   & 0.830  & \textbf{0.983}   & 0.924  & 0.775  & \textbf{0.994} & 0.947 & \multicolumn{1}{l|}{0.846} \\ \hline
\end{tabular}
\end{table}

\section{Discussion and Conclusion}

We successfully developed and validated a novel MNN based method for aneurysm segmentation and isolation and compared its performances to current state-of-the-art ACP method~\cite{jerman2019automated}. To our knowledge our method is the first that works on all aneurysm cases regardless of their own shape or shape of the surrounding vessels. At the same time, the ACP method fails to provide correct results in approximately 10\% of the cases and provide partially satisfactory segmentations on 30\% of our cases. The computational times, compared to approximately 11 minutes for the ACP, were less then 23 seconds for the proposed method. 

The proposed method achieved 15\% better median sensitivity than the ACP. For comparison we use the least-square minimization to fit a plane to the original output of our method and still achieved 10\% better median sensitivity than ACP. We succeeded to isolate aneurysm from vessel for all cases with the sensitivity of at least 0.759 (median sensitivity for all cases was 0.985, while for validation dataset the sensitivity was 0.994). The 5 validation cases were not used in training and were randomly chosen from the set of 100 aneurysms. 

Figure~\ref{results} shows 2 cases where all methods work and 2 cases where current state-of-the-art ACP method fails. Red curve is manual reference neck curve determined by skilled neurosurgeon, yellow/purple areas on meshes represents our isolation of aneurysm from vessel, black surface is a manually positioned cutting plane, while the orange surface represents the plane fit on border points between aneurysm and vessel of our method's output. In sections \textbf{A} and \textbf{B} of Figure~\ref{results} all approaches worked well, while in sections \textbf{C} the ACP method failed, while the plane acquired from output of our method still provided good results, in section \textbf{D} only our method provided meaningful results.   

We have successfully surpassed the sensitivity of MNN to rotation and scaling of input data with random rotations and scaling of training data. The additional rotation of target mesh contributed to better and more robust prediction of our MNN model. The output in the form of heatmap allows us to see the probability of each point being part of aneurysm or vessel. Probability map also emphasizes the area between aneurysm or vessel (the values smaller than one, but bigger than zero).

We have also shown that point cloud based approach is not sensitive to different input image modalities and artifacts associated to them. To create our data we used all common angiography modalities, namely the CTA, DSA and MRA. 

Accurate and fast prediction of aneurysm neck is essential for computing aneurysm features that are based on neck curve determination, such as volume, surface area, sphericity index, surface are to volume ratio, etc. Those measurements are useful in a variety of clinical and research applications related to intracranial aneurysms. An important treatment procedure called coiling involves coil embolization for treating unruptured aneurysms. The quantity of inserted coil depends on the measured aneurysm volume. Reliable volume calculations can prevent complications during and after the procedure, for instance, caused by inaccurate amount of inserted coil that can cause aneurysm lumen reopening and eventually result in rupture~\cite{slob2005relation,kai2005evaluation}.

Surface area and volume have both been demonstrated to be important indicators of aneurysm rupture risk \cite{austin1989controlled,valencia2008blood}. As shown on Figure \ref{results} the segmentation and isolation with cutting plane does not seem robust enough to isolate aneurysm from the vessels regardless of the position and the shape of aneurysm. Though prior work on isolation with non-planar separation curves showed some promising results \cite{saalfeld2018semiautomatic,lawonn2019geometric}, none of them achieved clinically acceptable results and robust enough performance. Small aneurysms with curved veins proved to be extremely challenging. Our method showed high accuracy and sensitivity, even with a relatively small dataset of 100 aneurysms. The method was able to predict, which points belong to the aneurysm automatically with sensitivity of 0.985. We feel that the robustness of our model will be reconfirmed  even on larger volumes of training and test data. 

The proposed deep learning approach to aneurysm isolation  can identify and label unordered aneurysm points in 3D vascular mesh automatically and with great sensitivity. This approach enables robust, repeatable and fast isolation of aneurysms from the surrounding vessels, thus the method seems readily usable in research and clinical applications. 
%
%
%
\bibliographystyle{splncs04}
\bibliography{mybibliography}

\end{document}